# A LONG-PERIOD, VIOLENTLY-VARIABLE X-RAY SOURCE IN A YOUNG SNR

A. De Luca[1*], P.A. Caraveo[1], S. Mereghetti[1], A. Tiengo[1], G.F. Bignami[2,3]

[1] INAF-IASF Milano, Via Bassini 15, I-20133 Milano, Italy
[2] CESR, CNRS-UPS, 9, Av. du Colonel Roche, 31028 Toulouse Cedex 4, France
[3] Università degli Studi di Pavia, Dip. di Fisica Nucleare e Teorica, Via Bassi 6, I-27100 Pavia, Italy
[*] to whom correspondence should be addressed; E-mail: deluca@iasf-milano.inaf.it

**Observations with the Newton X-ray Multimirror Mission (XMM) show a strong periodic modulation at 6.67±0.03 hours of the X-ray source at the centre of the 2,000-year-old supernova remnant RCW 103. No fast pulsations are visible. If genetically tied to the supernova remnant, the source could either be an X-ray binary, comprising a compact object and a low-mass star in an eccentric orbit, or an isolated neutron star. In the latter case, its age-period combination would point to a peculiar magnetar, dramatically slowed-down, possibly by a supernova debris disc. Both scenarios require non-standard assumptions on the formation and evolution of compact objects in supernova explosions.**

RCW 103 is a young (~2 ky), shell-type Supernova Remnant (SNR), with an X-ray point source very close to its centre. From its discovery (*1*), this source (1E161348-5055, hereinafter 1E), characterized by unpulsed, soft X-ray emission and no radio or optical counterpart (*2*), has been considered a candidate Neutron Star (NS), obviously young, assuming it was born in the same supernova event which originated the surrounding SNR. Subsequent X-ray observations have shown 1E to have a peculiar temporal behaviour, with orders-of-magnitude secular flux variations (*3,4*). Variabilities and tentative periodicities were proposed in the range from 1 to 6 hrs (*5,6,7*). At the SNR distance from Earth (~3.3 kpc), 1E has a luminosity of $10^{33}$-$10^{35}$ erg s$^{-1}$. Optical/IR observations point to an underluminous counterpart for this object: 3 faint IR objects, with H~22-23, lie in the Chandra error circle (*8*), but nothing is detected at visible wavelengths (R>25.6, I>25) (*6,9*).

Here we report results obtained during a long, uninterrupted XMM-Newton observation on 2005, August 23rd, using the positive-negative (pn) (*10*) and Metal-Oxide Semiconductor (MOS) (*11*) cameras of the European Photon Imaging Camera (EPIC) instrument. The total observing time was 89.2 ks. Details on data analysis are given in the Supporting Online Material. In the resulting RCW 103 MOS image (Fig.1) 1E stands out at less than 20" from the geometrical centre of the remnant, itself ~9' in diameter.

Our data show unambiguously that the source is periodic (Fig.1 B). The best estimate of the period is 6.67±0.03 hours (24.0±0.1 ks). The flux modulation is large, with a pulsed fraction PF=43.5±1.8%. A search for fast periodicities was performed, with negative results. Periodicities with P≥12 ms and pulsed fraction PF≥10% are excluded at the 99% confidence level.

As described in the SOM, time-averaged spectra for source and background were extracted for each EPIC detector in the 0.5-8 keV range. Single-component models do not yield acceptable results. The best fit is found for a double-component model consisting of a blackbody curve with temperature kT~0.5 keV and emitting radius of ~600 m, contributing ~70% of the flux, complemented either by a second blackbody (kT~1 keV) or by a steep power law (photon index $\Gamma$~3) (see Figure S1 and Table S1). The observed, time-averaged flux is $1.7\times10^{-12}$ erg cm$^{-2}$ s$^{-1}$ (0.5-8 keV).

The high XMM/Newton throughput and the long observing time yielded a total of 46900 photons, i.e. a good enough statistics to allow us to perform phase-resolved spectroscopy. We find evidence for a definite hardening (higher average photon energy) of the 0.5 to 8 keV spectrum at the light curve peaks and a softening at the troughs. Such a spectral evolution may be modelled by a higher temperature and larger emitting area of the dominant blackbody at the peak, coupled to a higher line-of-sight absorption (see Fig.S2).

The long-term time behaviour of 1E prior to our 2005 observation was also studied and is summarized in Figure 2. In Aug. '05, the source was clearly caught in a low state. We re-analyzed the 50 ksec 2001 XMM data, when the source was in a higher state (~$10^{-11}$ erg cm$^{-2}$ s$^{-1}$). The periodicity seen in 2005 can be recognized (Fig.2 B, upper curve), albeit with a smaller pulsed fraction (PF=11.7±1.4%). The period extracted from the 2001 data is 6.72±0.08 hours (24.2±0.3 ks), consistent with the 2005 value, with no evidence for a period derivative. However, the source phenomenology is completely different. Apart from a factor of ~6 difference in the average flux value, the 2001 light curve has a much more complex substructure. We note that the pulsed flux in 2001 is similar to the 2005 one, or ~$2\times10^{-12}$ erg cm$^{-2}$ s$^{-1}$. The source time-averaged spectrum is significantly harder than in 2005, with a larger contribution from the high-energy component, as well as a larger absorption (Fig.S3 and Table S1).

The 6.67 hrs periodicity reported here, as well as the long-term flux variability and complex spectral behaviour, render 1E unique among young compact objects still embedded in their SNR and make any interpretation very difficult.
The association of 1E and RCW103 appears very robust, based on their perfect positional coincidence as well as on radio studies suggesting consistent distance estimates for the two objects (*12*). The chance alignment of a foreground object with the centre of RCW103, can be excluded on the basis of the optical data. While an AM Her system (*13*) at 50-100 pc could show an X-ray phenomenology somewhat similar to the observed one, it would imply a ~10 magnitudes brighter optical/IR counterpart.
Thus, we will assume that 1E was born together with its host SNR, which is 2,000 yr old (*14*).

Interpreting the 6.67 hr as an orbital period, we first explore a binary system hypothesis for 1E, featuring a compact object (either a NS or a black hole) born in the SN explosion, and a faint star, for which existing optical/IR data (*6,8,9*) set stringent constraints. The colours and luminosity of the possible counterparts, assuming an interstellar reddening $A_V$~4.5 (*12*), are only compatible with a M-class dwarf of ~0.4$M_\odot$ or smaller. 1E would thus be a Low-Mass X-ray Binary (LMXB) which survived the SN event. However, 1E's phenomenology is very unusual for a LMXB. Its highly variable X-ray luminosity (~$10^{33}$-$10^{35}$ erg s$^{-1}$) is low, both compared to the peak luminosities of transient LMXBs (~$10^{38}$ erg s$^{-1}$) and to the persistent LMXB output ($10^{36}$-$10^{37}$ erg s$^{-1}$). It is a luminosity similar to that of Very Faint X-Ray Transients (*15*), which are, however, very old systems ($10^9$ y).
Moreover, the pronounced orbital modulation and spectral phase variability reported here have never been observed in LMXBs. The same is true for the 1E long-term evolution, with its dramatic orbital modulation change and long outburst decay (Fig.2).

Young age could be the explanation. Standard LMXBs are at least hundreds of millions of years older than 1E and have had enough time to evolve (*16*) to a phase in which the donor star, having filled its Roche lobe, supplies a large mass transfer towards the compact object, via an accretion disc, at a rate close to the Eddington limit. Conversely, in a very young, very low-mass binary, survivor of a supernova explosion, a significant orbital eccentricity is expected (*17*), with an important role in controlling any mass transfer within the system. For a dwarf star mass $M_d$ in the range of 0.2-0.4 $M_\odot$, an orbit eccentricity $e$ of 0.5-0.2 would position the L1 point just above the dwarf star surface at periastron. Mass-exchange will thus become possible within a narrow range of orbital phases, with a transit time of ~10 minutes from the donor through L1 towards the compact object. The transferred material, with its large angular momentum, will start settling in a disc. One also expects significant orbital modulation in the fraction $f$ of the dwarf star wind mass captured by the compact object. In the Bondi-Hoyle approach (*18*), $f \propto d^{-2} v_{rel}^{-4}$, where $d$ is the orbital separation and $v_{rel}$ is the relative velocity between the dwarf star wind and the compact object. In 1E, $f$ would vary by a factor ranging from 2.5 to 9 for an $M_d$ of 0.4-0.2 $M_\odot$ and an $e$ of 0.2-0.5, with a single peak during the descending part of the orbit, where the combination of velocities is most favourable (Fig.S4).

A "double accretion" scenario could thus be at work: wind accretion provides the bulk of the sharply-peaked 6.67 hr modulation, while a disc controls 1E's long-term variations. Flux outbursts could be due to episodic mass ejections from the dwarf star and/or by disc instabilities, while dips in the light curve could be due to occultations by disc structures.

X-ray production remains the crucial point in such a binary model. If the compact object is a NS, accretion can occur if both the rotating dipole ("ejector") and the centrifugal ("propeller") barriers (*16*) can be overcome: the dynamical pressure of the infalling gas must exceed the pressure of the NS dipole radiation, at least down to a distance where corotation with the NS is slower than the keplerian velocity. This would imply the 2 ky-old NS to have a very low magnetic field and/or a slow rotation period. Indeed, to produce via accretion a luminosity in the range observed for 1E, the above conditions imply (*19*) $P \sim [0.35-2.5] B_{10}^{6/7}$ s, where $B_{10}$ is the B-field in units of $10^{10}$ G. These values are very peculiar if compared with the canonical picture of a standard 2kyr old NS, having a few $10^{12}$ G B-field and spinning at a few tens of ms. On the other hand, if a black hole is present, accretion processes at the low rates implied by our young binary scenario are expected to be highly inefficient (*20*), and production of the observed luminosities could be problematic.

Faced with a highly non-standard binary picture, we will also consider an isolated-object scenario for 1E. We will focus on NSs, since the periodical modulation rules against a black hole. The 6.67 hr periodicity could be related to the free precession of a fast-rotating NS, with the X-rays coming from a surface hot spot modulated at the precession period (*21*). However, we find no trace of the expected, faster periodicity related to the star rotation. We cannot exclude a peculiar emission geometry, somewhat symmetrical with respect to the rotation axis, but find it an unlikely possibility. Similarly, a NS rotation period shorter than 12 ms seems unlikely, since some evidence of a synchrotron nebula, or of non-thermal emission, due to the rotating dipole radiation, would be felt. A non-thermal X-ray output from ~$5 \times 10^{34}$ erg s$^{-1}$ to ~$5 \times 10^{36}$ erg s$^{-1}$ is expected (*22*) for a 12 ms pulsar with a $10^{12}$ G B-field. Moreover, any precession scenario would not explain the dramatic flux outbursts, together with the other long-term changes in the source phenomenology (Fig.2).

Alternatively, 1E could be a normal isolated NS, slowly rotating at the 6.67 hr period. Some huge braking mechanism would have to be invoked to slow it down over 2,000 yrs from its presumably much shorter birth period. To do this with the classical dipole-radiation pulsar mechanism requires the unheard of, and probably unphysical, magnetic field value of B~$10^{18}$ G. On the other side, even if 1E were a "normal" NS with a birth period close to 6.67 hrs, this would not account for its long-term X-ray flux variability.

1E could, instead, be a magnetar, a neutron star with an ultra-high magnetic field of order $10^{15}$ G (*23*), now rotating at 6.67 hrs. Indeed, all types of X-ray variabilities observed for 1E, as well as its luminosity and spectral shape, would be naturally explained in the magnetar frame. Magnetar candidates (namely, Anomalous X-ray Pulsars – AXPs - and Soft Gamma Repeaters - SGRs) show long-term variations in flux, spectrum, pulse shape and pulsed fraction. All magnetars, however, spin more than 1,000 times faster than 1E, with periods well clustered in the 5-12 s range. The slowing-down mechanism obviously required for 1E could result from the transfer, through its rotating giant B-field, of the star's angular momentum to the material of a hypothetical SN debris disc ("propeller" effect). Our detailed calculations (*24, 25*) show that a disc of $3\times10^{-5}$ M$_\odot$ would have been enough to slow down, over 2,000 yrs, a B=$5\times10^{15}$ G magnetar, provided it was born with P$\gtrsim$300 ms (Fig.S5). Such a birth period is necessary for avoiding an early "ejector" phase, since the pressure of the radiation of the rotating dipole quickly pushes away any disc surrounding a fast magnetar. With a slower rotation at birth, the star instead retains its disc and starts immediately a very efficient rotational energy loss. A birth period $\gtrsim$300 ms is too long to fit in the most popular explanation (*26*) for the origin of the huge B-fields of magnetars ("dynamo effect" in the proto-neutron star, requiring P$_0$~1 ms). However, alternative high B-field formation scenarios (e.g. compression of the progenitor field) have been proposed (*27, 28*), based on possible evidences that not all magnetars are born as very fast rotators (*27*).

The recent discovery of a debris disk around an AXP (*29*) may support a "braked magnetar" picture for 1E, suggesting that at least some magnetars could be surrounded by fossil disks. AXPs and SGRs, as witnessed by their 5-12 s periods, did not experience an efficient propeller phase, possibly due to a shorter period at birth or strong gamma-ray bursting activity. If 1E is indeed a "slow magnetar", this implies a totally new evolutionary channel for isolated NSs, one in which their spin history is dominated by SN debris. The fraction of NSs following such a channel should be small, however, considering the uniqueness of 1E among compact object associated with SNRs. Furthermore, as for standard AXPs and SGRs, it may be that such objects rapidly ($10^5$ yr ?) become unobservable. However, one could think of some compact objects not showing now any pulsation, in spite of large observational efforts, such as sources associated with young SNRs (e.g. Cas A, VelaJr, G347.3-0.4) (*6*).

Other pictures could also be explored. We may consider a peculiar binary system in which the 6.67 hr periodicity reflects the spin period of the collapsed object (necessarily a NS), but in which the orbital period is much longer and undetected. As in the isolated NS case, the main difficulty here is to account for a huge braking of the NS rotation in 2000 yr, unless it was born spinning at 6.67 hrs. As in the case of 2S 0114+650 (*30*), a binary system featuring a 2.7 hr period NS and a giant companion with an estimated age of > $10^7$ y, the only viable mechanism could be the propeller effect on the companion star wind. Following (*19, 30*), we note that for plausible parameters of the accretion rate in the 1E system (Mdot~$10^9$-$10^{10}$ g s$^{-1}$), an equilibrium period of 6.67 hrs could indeed result for NS B-fields of order $10^{12}$-$10^{13}$ G, but the overall NS spin-down process would require $10^8$-$10^{10}$ yr. Assuming instead a magnetar B-field of $5\times10^{15}$ G and a higher (but still plausible) accretion rate of $10^{13}$ g s$^{-1}$, the braking would be much more efficient, but in any case > 20 kyr would be required to reach a period similar to the observed one. Thus, such a picture seems untenable. One could invoke that 1E and RCW103 were generated by 2 different SN explosions within a same binary system, originally composed of two high-mass stars. The first SN produced 1E (at least ~$10^5$ yr ago, to allow for the fading of the resulting SNR) and did not disrupt the binary. The second produced RCW103 ~2,000 yr ago, but did not leave any visible compact object. 1E could have been slowed down over the life time of its companion star (~$10^7$ yr?), remaining in any case an active magnetar, as is required to explain its time behaviour. Okham's razor argues against such an interpretation: for a scenario involving a magnetar, the braking of a young isolated object by SN debris seems the most plausible explanation.

Many more details remain to be explored for both 1E and RCW 103. Deeper and longer X-ray observations could detect fast pulsations, ruling out the slow rotator model proposed above. Observations during the source "high state" could allow for phase-resolved spectroscopy, giving evidence for intervening circumstellar occulting material. Optical/IR observations could yield the nature of any optical counterpart, to check, for example, on the presence of a disc. It would be also useful to carry out spectral studies of the diffuse remnant material. Although difficult, such studies could be crucial in understanding a SN event which, 2,000 years ago, created either a compact object or a binary system unique in its physical properties.

31. The XMM-Newton data analysis is supported by the Italian Space Agency (ASI) through contract ASI/INAF I/023/05/0. ADL and AT acknowledge an ASI fellowship

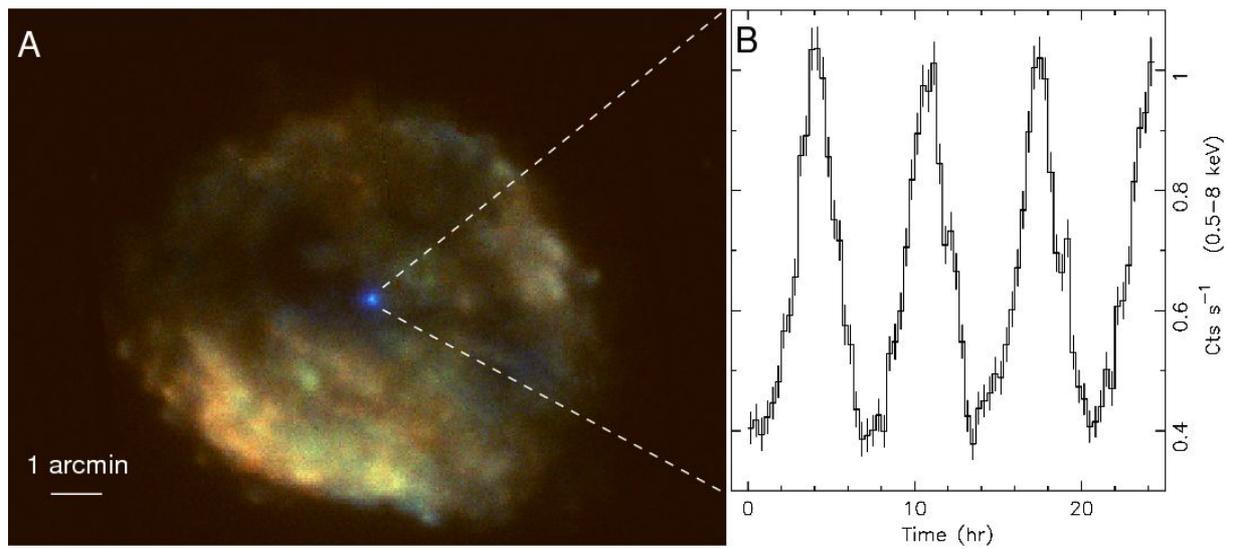

**Figure 1. (A)** The young supernova remnant RCW103 and its central source 1E161348-5055 (1E) as observed in August 2005 by the EPIC/MOS cameras onboard XMM-Newton. Photon energy is color-coded: red corresponds to the energy range 0.5-0.9 keV, green to 0.9-1.7 keV and blue to 1.7-8 keV. North is up, East is left. **(B)** Background-subtracted flux evolution of 1E in the 0.5-8 keV energy range, with its unambiguous 6.67 hr periodicity.

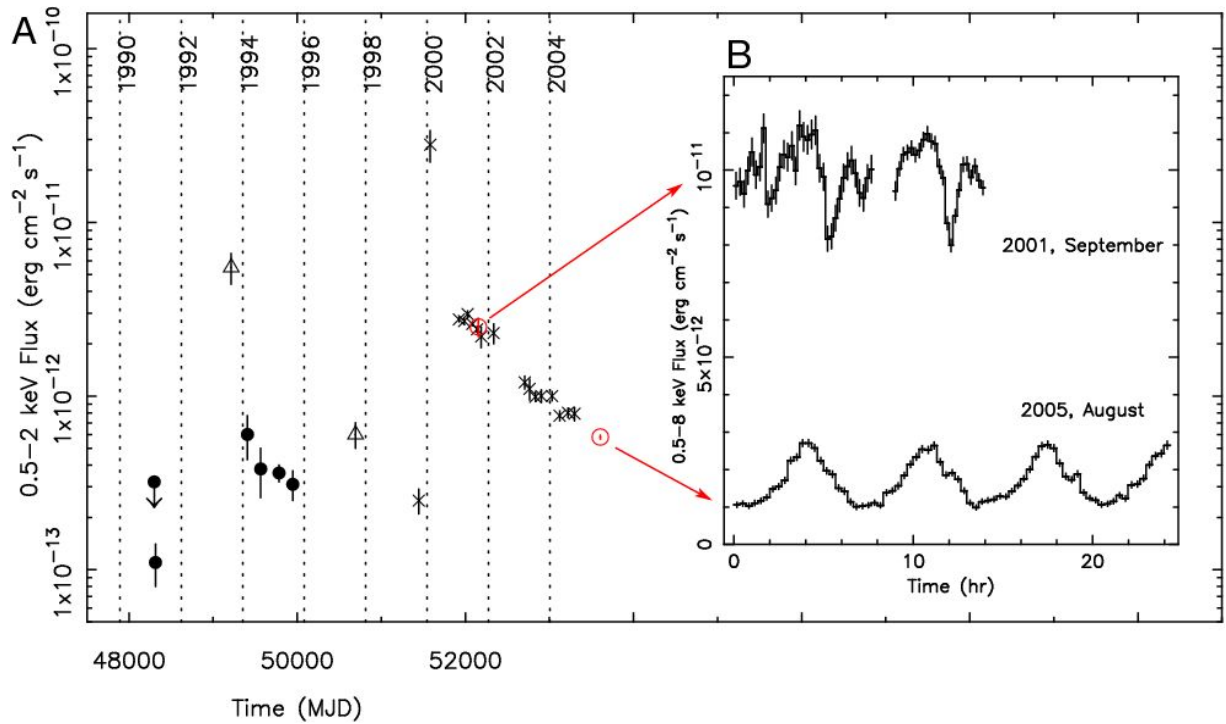

**Figure 2.** A synoptic view of time variabilities of 1E. **(A)** Source secular flux evolution as derived from our analysis of the public Chandra observations (crosses) performed between 1999 and 2005 and two XMM-Newton ones (red empty circles) performed in 2001 and 2005. A large, two-order of magnitude outburst between 1999-2000 is followed by a continuous fading down to the level of our 2005 XMM observation. Historical measurements (*3*) with the Rosat (black filled circles) and ASCA (triangles) satellites are also included and show another episode of flux increase around the ASCA observation in 1993. Source outbursts could thus be recurrent on a several-year timescale. **(B)** source flux variation over the '01 (upper curve) and '05 (lower curve) XMM observations, of respectively 50 ks and 90 ks. Observation starting times have been aligned, but no folding has been performed. The 6.67 hr periodicity may be now recognized in the shorter 2001 observation, although with a smaller pulsed fraction and with a much more complex pulse shape.

## Supporting online material

*1. XMM-Newton/EPIC data analysis*

*Observations and data reduction.* The long 2005 XMM-Newton observation of 1E started on August 23$^{rd}$ 2005 and lasted for 89.2 ks. The EPIC/pn was operated in Small Window mode with the thin optical filter; the EPIC/MOS1 and MOS2 were operated in Full Frame mode with the medium optical filter. Raw data were processed with the most recent release of the XMM-Newton Science Analysis Software (SASv6.5.0) using standard pipelines. Dead-time corrected exposure times were of 86.1 ks, 86.2 ks and 61.2 ks in the MOS1, MOS2 and pn detector, respectively. No time filtering to screen high particle background episodes was applied. Standard event PATTERN filtering was used (PATTERN 0-4 and 0-12 for the pn and MOS, respectively). Owing to the bright and patchy emission from the surrounding SNR, selection of source and background events required a particular care. We extracted background counts from a 20" radius circle centered at RA 16:17:42.4, Dec -51:02:38, where the surface brightness is comparable to the one seen in the source region (as estimated using high-resolution Chandra images). Selecting source events from a 15" radius circle yields background-subtracted count rates of 0.382±0.003 cts s$^{-1}$, 0.132±0.001 cts s$^{-1}$, 0.139±0.001 cts s$^{-1}$ in the 0.5-8 keV energy range for the pn, MOS1 and MOS2 in 2005. The same event selection was used to extract both light curves and spectra. Different choices of background regions result in somewhat different best fit values for the spectral parameters (mainly the $N_H$), but do not affect the source variability analysis.

*Time-integrated spectroscopy.* Time-integrated spectra for source and background were extracted for each EPIC instrument from the regions described above. Source spectra were rebinned in order to have at least 30 counts per spectral bin. Ad-hoc response matrices and effective area files were generated using SAS tasks rmfgen and arfgen. Spectral analysis was performed with the XSPEC v11.3 software, fitting simultaneously pn and MOS spectra in the 0.5-8 keV energy range. Single component models (including interstellar absorption) did not yield satisfactory results: while a power law is clearly inadequate (reduced $\chi^2 > 2.2$, 492 d.o.f.), a blackbody or a thermal bremsstrahlung curve yield somewhat better fits (reduced $\chi^2$ of 1.43 and 1.37, respectively, 492 d.o.f.), although inspection of the residuals shows significant, structured deviations. Better results were obtained using double component models consisting of a dominant blackbody curve complemented with a second blackbody (reduced $\chi^2=1.18$, 490 d.o.f.) or with a power law (reduced $\chi^2=1.19$, 490 d.o.f.). The best fit parameters are reported in table S1. The observed spectra together with the best fit double blackbody model are shown in Fig.S1, where residuals are also plotted in the lower panel.

*Timing analysis.* Source and background light curves were extracted for each instrument using the regions described above. The source periodicity was immediately apparent. Since the light curve is sharply peaked, it cannot be satisfactory fitted using a sin function. Indeed, while a sin function fit to the combined (pn+MOS) background-subtracted light curve (0.5-8 keV) yields a reduced $\chi^2=2.67$ (84 d.o.f.), adding a second harmonic improves the overall fit quality (reduced $\chi^2=1.63$, 82 d.o.f.). However, the resulting best fit period of P=6.67±0.03 hr remains identical. MOS and pn light curves were also independently analysed, yielding fully consistent results. The Pulsed Fraction, defined as PF=(CR$_{max}$-CR$_{min}$)/(CR$_{max}$+CR$_{min}$), where CR$_{max}$ and CR$_{min}$ are the observed background-subtracted count rates at the peak and at the minimum, is very large: in the 0.5-8 keV range the PF turns out to be 43.5±1.8%. Although the pulse profile does not change significantly as a function of energy (Fig.S2), there are hints for substructures in the soft band (0.5-2 keV) but not in the hard one (2-8 keV). The pulsed fraction increases significantly as a function of energy, going from 37.1±2.8% in the soft band to 56.8±2.9% in the hard band. The corresponding hardness ratio (Fig.S1), defined as HR=(CR$_{Hard}$-CR$_{Soft}$)/(CR$_{Hard}$+CR$_{Soft}$), where CR$_{Hard}$ and CR$_{Soft}$ are the background-subtracted count rates in the 2-8 keV and in th 0.5-2 keV ranges, shows that the source emission is markedly harder at the peak and softer at minimum, hinting a significant spectral evolution as a function of the 6.67 hr cycle. A search for fast periodicity was also performed, using standard fast Fourier transform techniques. Since our high resolution timing data have a ~6 ms sampling, we searched for periodicities larger than 12 ms. No pulsed signal was found. Assuming a sinusoidal pulse shape, we constrain the source pulsed fraction to be < 10% (99% confidence level ).

*Phase-resolved spectroscopy.* In order to study the spectral evolution as a function of the 6.67 hr cycle, we extracted source and background spectra selecting only events within the phase intervals corresponding to the peak and to the minimum of the light curve (Fig.S2 A). The spectral analysis was performed assuming the double blackbody model best fitting the time-integrated spectrum. As a first step, we kept all parameters fixed to their best fit values for the time-integrated spectrum, allowing only the overall normalization factor to vary. This approach yielded unacceptable fits to the data (reduced $\chi^2$=3.75, 632 d.o.f.), with residuals clearly showing the source spectrum to be softer at minimum and harder at the peak. A better fit was obtained allowing the normalizations of the two spectral components to vary independently (reduced $\chi^2$ ~ 1.4, 628 d.o.f.); a further improvement (reduced $\chi^2$=1.15, 624 d.o.f.) was achieved allowing the blackbody temperatures to vary. Such best fit parameters are reported in the caption to Fig.S2. The spectra obtained for the peak and the minimum are shown in the upper panel of Fig.S2 (B), while the lower panel gives the ratio between the peak spectrum and the best fit model for the spectrum of the minimum (renormalized to the same number of counts). The spectral hardening at the peak is apparent.

## The September 2001 observations.

*Observations and data reduction.* The September 2001 dataset results from the merging of two observations. The first pointing (30 ks) was aimed at the energetic pulsar PSR J1617-5055 (located ~7.5 arcmin North of RCW103) and useful data on 1E (imaged at 7.5 arcmin offaxis) were collected only by the MOS cameras (operated in Full Frame mode with the medium optical filter). The satellite then slewed to RCW103 and collected 20 ks of data on 1E using all EPIC cameras, operated with a setup identical to that used in 2005. Data reduction and analysis was performed as for the 2005 observation. Dead-time corrected exposure time was of 27.8 ks (off-axis observation) + 18.8 ks (on-axis observation) and 27.8+18.9 ks for MOS1 and MOS2 and 12.9 ks for the pn. The background-subtracted on-axis count rate of 1E was of 1.936±0.013 cts s$^{-1}$, 0.692±0.006 cts s$^{-1}$ and 0.701±0.006 cts s$^{-1}$ for the pn, MOS1 and MOS2, respectively, i.e. a factor ~5 higher than in 2005.

*Time-integrated spectroscopy.* The spectral analysis was performed following the procedure described for the 2005 observation. Two component models, either encompassing two blackbody curves (reduced $\chi^2$=1.09, 517 d.o.f.), or the combination of a blackbody curve and a power law (reduced $\chi^2$==1.14, 517 d.o.f.) yielded the best fit to the data, as reported in Table S1. The time-averaged observed flux is 9.9×10$^{-12}$ erg cm$^{-2}$ s$^{-1}$ (0.5-8 keV), i.e. ~6 times higher than in 2005. As shown in Fig.S3, the spectrum of the source in its 2001 "high state" is remarkably harder than in the "low state" of 2005. To reproduce the 2001-2005 spectral change, both spectral components in both models have to vary.

*Timing analysis.* Background-subtracted light curves were produced following the procedure described for the 2005 observation. In order to combine the data collected in the two (off-axis and on-axis) observations to produce a unique, 50 ks long light curve, we computed count-rate to flux conversion factors accounting for the observed source spectrum (see above) as well as for the significant vignetting affecting the first section of the data. The resulting light curve (0.5-8 keV energy range) is shown in Fig.2. With a light curve covering ~2 periods, the source periodicity may be clearly recognized, in spite of the 3 ks interruption. The pulse profile is very complex, with a narrow dip, ~2500 s long, occurring after the highest peak and a secondary, less pronounced, dip, separated by 0.5 in phase from the first one. The central times of such dips (computed by fitting a gaussian curve to their profile) were used to evaluate the period, which turned out to be of 6.72±0.08 hr, consistent with the value measured in 2005. The pulsed fraction is PF=11.7±1.4%, remarkably lower than in 2005. We note that, in absolute units, the pulsed flux is very similar to the value observed in 2005, i.e. ~$2\times10^{-12}$ erg cm$^{-2}$ s$^{-1}$.

*Time-resolved spectroscopy.* In order to investigate the spectral variation along the periodic cycle, we extracted time-resolved spectra selecting two 2500 s time intervals centered on the lowest dip and on to the highest peak, respectively. The time-integrated double blackbody best fit model was assumed as a template. A simple renormalization of such model does not fit well the data, since the dip spectrum is markedly harder. The spectral evolution may be described (reduced $\chi^2$ =1.02, 348 d.o.f.) by a smaller emitting radius of the dominant blackbody, possibly coupled to extra absorption along the line of sight, during the dip. However, a detailed modelling is hampered by the limited statistics.

## *2. Chandra/Advanced CCD Imaging Spectrometer (ACIS) data analysis.*

``Level 1" event files were retrieved from the public Chandra archive and processed with standard pipelines (acis_process_events, *S1*) using the Chandra Interactive Analysis of Observations (CIAO v3.2.1) software package. Source events were selected from circular regions corresponding to encircled energy fractions of ~0.9. In February 2000 the source was in a very high state with the data heavily affected by pile-up. To overcome the problem, we excluded the PSF core, extracting source events from an annulus with inner and outer radii of 2.5" and 6", respectively, and we recomputed the total flux evaluating the encircled energy fraction in the annular region (a conservative 20% error on the flux was assumed in such case). Background events were selected in all cases from 20" radius regions located close to the target. We extracted spectra for source and background and we generated appropriate response matrices and effective area files using the psextract (*S1*) script. Source spectra were rebinned in order to have at least 30 counts per channel. Each spectrum was fitted in XSPEC v11.3 in the 0.8-8 keV energy range using a double blackbody curve modified by interstellar absorption. The observed flux in the 0.5-2 keV range was then computed for each epoch using the corresponding best fit model.

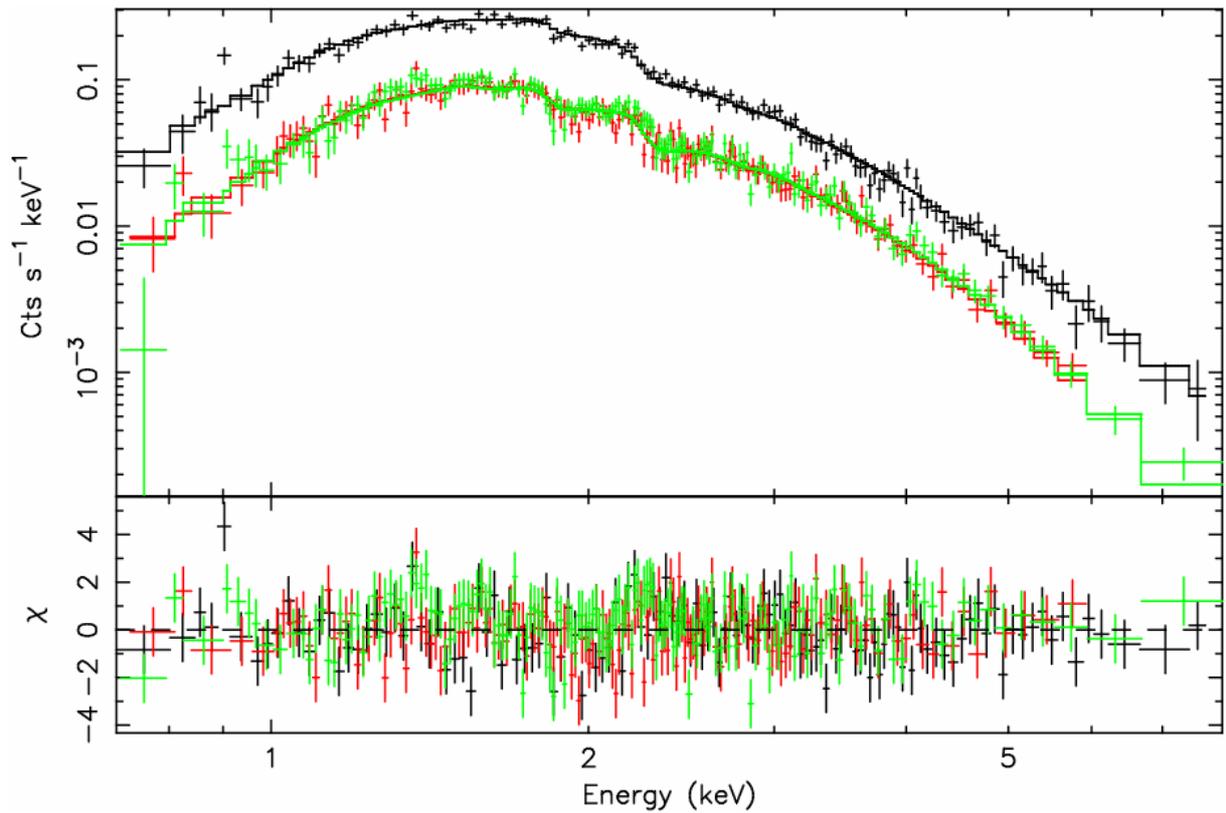

**Figure S1.** August 2005 spectrum of 1E as observed by the three EPIC cameras. The upper panel shows the time-integrated spectrum of 1E as observed with EPIC. Black, green and red data points are from the pn, MOS1 and MOS2 cameras, respectively. The best fit model, consisting of the sum of two blackbody curves (see text), is overplotted. The lower panel shows the residuals in units of statistical errors.

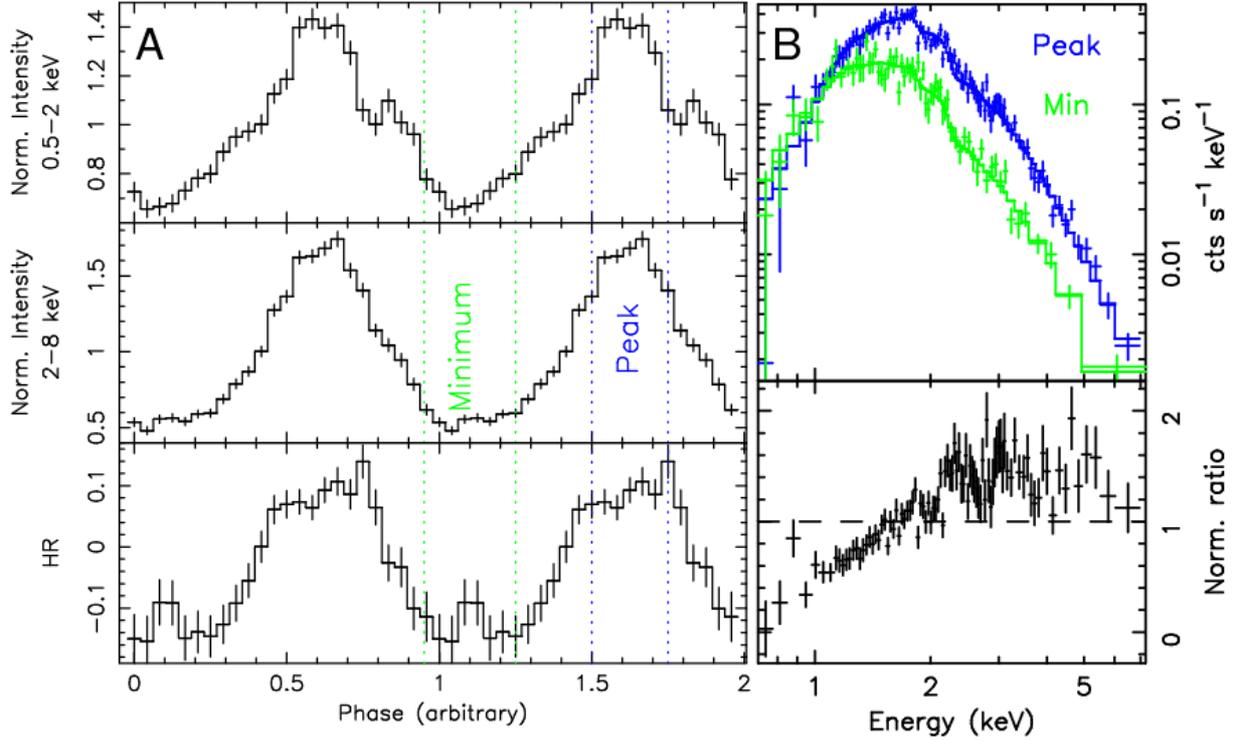

**Figure S2.** Spectral evolution of 1E over its 6.67 hours periodicity, as observed in August 2005. **(A)** Folded light curves (normalized to the average flux value) in the soft (0.5-2 keV) and in the hard (2-8 keV) ranges and corresponding hardness ratio, defined as HR=(Cts$_{2-8}$-Cts$_{0.5-2}$)/(Cts$_{2-8}$+Cts$_{0.5-2}$). Note the spectral hardening during the pulse peak, as well as the energy dependence of the pulsed fraction (PF=37.1±2.8% in the soft band, PF=56.8±2.9% in the hard band). **(B)** Spectra of 1E as extracted from the phase intervals corresponding to the peak (blue) and to the minimum (green), as well as the ratio between the peak spectrum and the renormalized best fit model for the minimum. The spectral evolution is quite complex: assuming as a template the double blackbody model, we obtain at minimum $N_H=5\pm1\times10^{21}$ cm$^{-2}$; temperature and emitting radius of the dominant blackbody $kT_{bb1}=0.47\pm0.03$ and $R_{bb1}=560\pm60$ m; temperature and emitting radius of the second blackbody $kT_{bb2}=1.4\pm0.7$ keV, $R_{bb2}<80$ m; at the peak $N_H=9\pm1\times10^{21}$ cm$^{-2}$, $kT_{bb1}=0.54\pm0.02$, $R_{bb1}=750\pm40$ m, $kT_{bb2}=1.3\pm0.4$ keV, $R_{bb2}=40\pm20$ m (errors are at 90% confidence level for a single parameter).

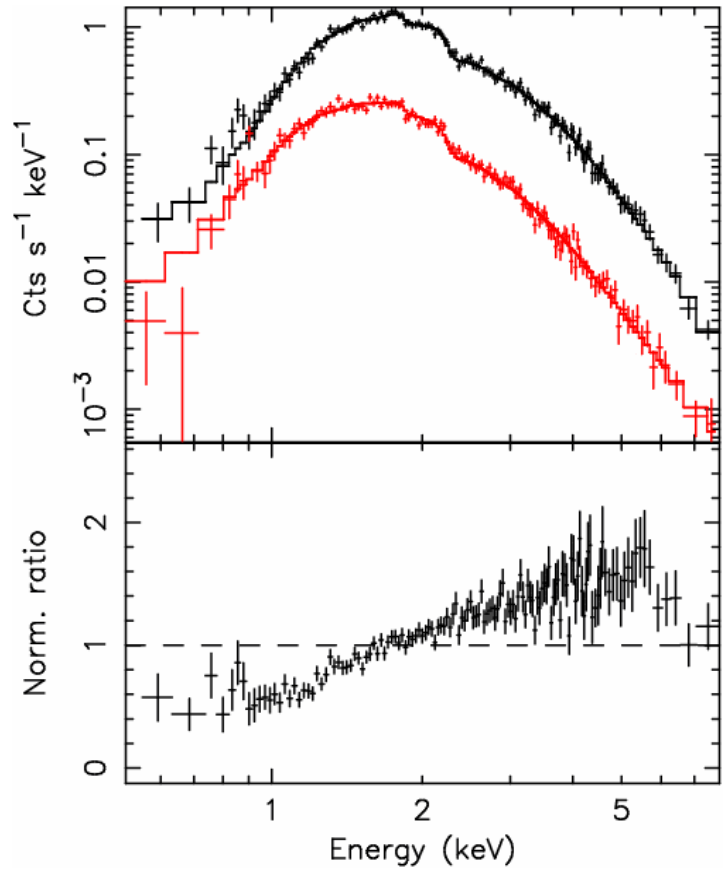

**Figure S3.** The upper panel shows the time-integrated pn spectrum of 1E in the high state of 2001 (black data points) together with the best fit double blackbody model (black line); the 2005 pn spectrum (red data point) and best fit model (red line) are also plotted. The lower panel shows the ratio between the 2001 spectrum and the 2005 best fit model (renormalized to the 2001 number of counts). The 2001 spectrum is clearly harder than the 2005 one.

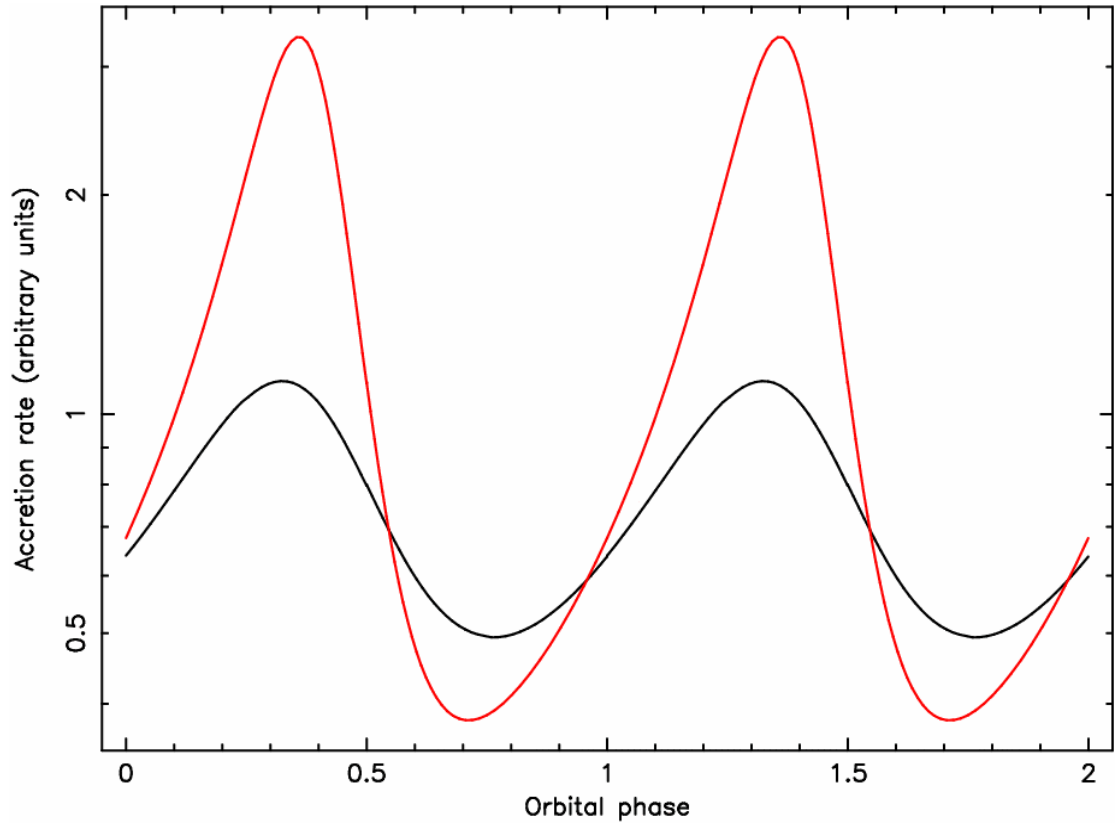

**Figure S4.** Expected modulation of the mass fraction *f* of the companion star wind captured by the accreting object in an eccentric system. We followed the Bondi-Hoyle approach, assuming a 1.4 $M_\odot$ neutron star and a 6.67 hr orbital period. The black line corresponds to the case of a 0.4 $M_\odot$ companion star and an eccentricity of 0.2; the red line corresponds to the case of a 0.2 $M_\odot$ companion star and an eccentricity of 0.5. A companion wind velocity of 300 km s$^{-1}$ was assumed. Phase 0 correspond to the periastron. A remarkable modulation is obtained in both cases, with a single peak in the descending part of the orbit. The resulting shape is very similar to the observed light curve of 1E in its low state.

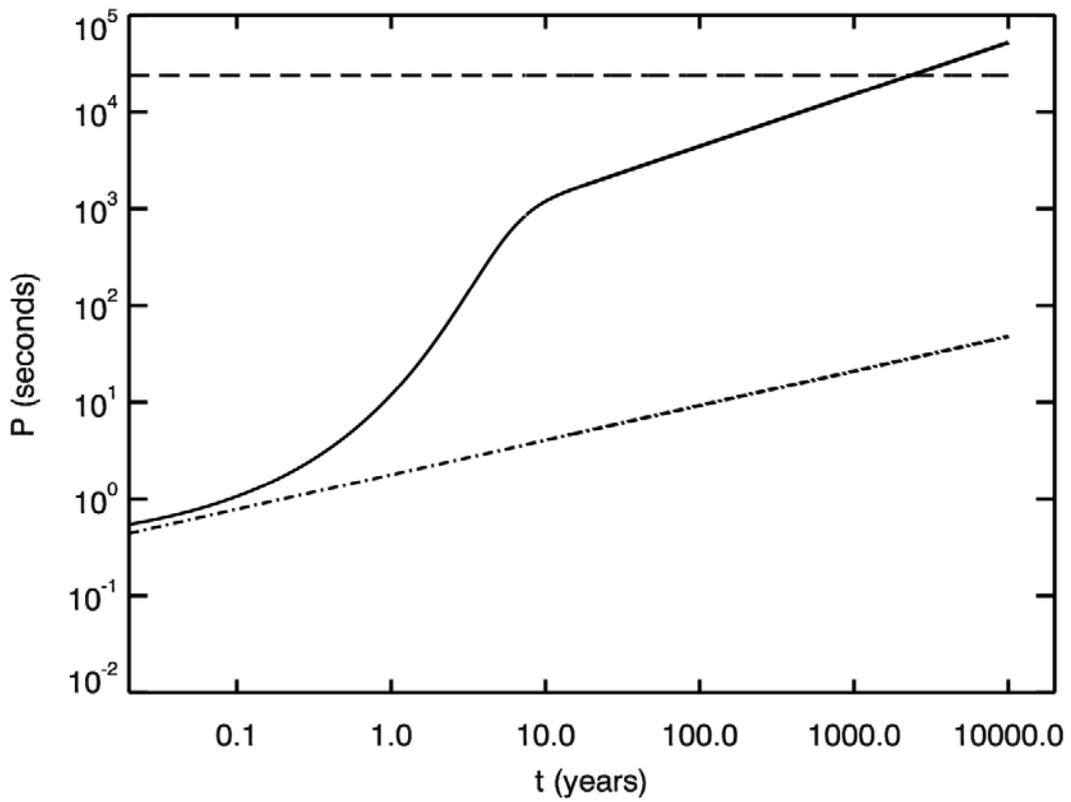

**Figure S5.** Spin-down history of 1E, resulting from propeller interaction with the material of a SN debris disc. The source spin period as a function of time is plotted as a black solid line. The dot-dashed line shows the faster spin period at which propeller interaction may occur. For faster rotation rates the pressure of the rotating dipole would exceed the pressure of the infalling material, pushing the disc outside the light cylinder (ejector phase), preventing the propeller effect. Following (*24, 25*), assuming a magnetic field of $5\times10^{15}$ G, a disc mass of $3\times10^{-5}$ M$_\odot$ and a birth period of 0.3 s, an early ejector phase is avoided and the NS starts spinning down in the propeller regime. In ~2 kyr the NS rotation is quenched to P~6.67 hr (~24 ks), marked by the horizontal dashed line.

**Table S1.** Results of XMM-Newton/EPIC time-integrated spectroscopy.

| Year | 2005 | 2005 | 2001 | 2001 |
|---|---|---|---|---|
| Model | BB+BB | BB+PL | BB+BB | BB+PL |
| $N_H$ ($10^{22}$ cm$^{-2}$) | 0.65±0.04 | 0.85±0.20 | 1.08±0.05 | 1.35±0.15 |
| $kT_{bb1}$ (keV) | 0.51±0.01 | 0.52±0.01 | 0.50±0.02 | 0.54±0.01 |
| $R_{bb1}$ (m) | 610±35 | 570±30 | 1560±120 | 1310±50 |
| $kT_{bb2}$ (keV) | $1.0^{+0.4}_{-0.2}$ | - | $0.93^{+0.13}_{-0.09}$ | - |
| $R_{bb2}$ (m) | $35^{+22}_{-12}$ | - | $220^{+100}_{-60}$ | - |
| $\Gamma$ | - | $2.9^{+0.4}_{-0.9}$ | - | $3.0^{+0.3}_{-0.5}$ |
| $F_{obs}^{a}$ (erg cm$^{-2}$ s$^{-1}$) | $1.7\times10^{-12}$ | $1.7\times10^{-12}$ | $9.9\times10^{-12}$ | $9.9\times10^{-12}$ |
| $L_{bb1}^{b}$ (erg s$^{-1}$) | $3.0\times10^{33}$ | $2.9\times10^{33}$ | $2.0\times10^{34}$ | $1.9\times10^{34}$ |
| $L_{bb2}^{c}$ (erg s$^{-1}$) | $3.8\times10^{32}$ | - | $4.6\times10^{33}$ | - |
| $L_{PL}^{d}$ (erg s$^{-1}$) | - | $1.6\times10^{33}$ | - | $1.7\times10^{34}$ |
| $\chi^2_\nu$ | 1.18 | 1.19 | 1.09 | 1.14 |
| d.o.f. | 490 | 490 | 517 | 517 |

[a] Observed Flux, 0.5-8 keV

[b] Bolometric luminosity of dominant blackbody

[c] Bolometric luminosity of second blackbody

[d] Luminosity of the power law component, 0.5-8 keV

**References and notes.**

S1. See http://cxc.harvard.edu/ciao/threads/all.html